\documentclass[aps,prb,twocolumn,groupedaddress,amsmath,
amssymb,amsfonts]{revtex4}

\bibdata{cont}

\DeclareMathOperator{\T}{T}

\DeclareMathOperator{\sgn}{sgn}
\DeclareMathOperator{\diag}{diag}
\usepackage{graphics}

\begin{document}

\title{Identifying the maximum entropy method as a special limit\\
of stochastic analytic continuation}

\author{K. S. D. Beach}
\email[]{ksdb@mit.edu}
\homepage[]{web.mit.edu/ksdb/www}
\affiliation{Department of Physics, Massachusetts Institute of
Technology}

\date{
March 1, 2004
}

\begin{abstract}
The maximum entropy method is shown to be a special limit of the
stochastic analytic continuation method introduced by Sandvik [Phys.\
Rev.\ B {\bf 57}, 10287 (1998)]. We employ a mapping between the
analytic continuation problem and a system of interacting classical
fields. The Hamiltonian of this system is chosen such that the
determination of its ground state field configuration corresponds to an
unregularized inversion of the analytic continuation input data. The
regularization is effected by performing a thermal average over the
field configurations at a small fictitious temperature using Monte Carlo
sampling. We prove that the maximum entropy method, the currently
accepted state of the art, is simply the mean field limit of this fully
dynamical procedure. We also describe a technical innovation: we suggest
that a parallel tempering algorithm leads to better traversal of the
phase space and makes it easy to identify the critical value of the
regularization temperature.
\end{abstract}


\maketitle

\section{Introduction}

Wick rotation transforms imaginary time correlation functions into real,
measurable response functions.  Analytical results, or numerical results
fit to a known functional form, allow for a simple substitution of
variables: \emph{e.g.}, $-i\tau \mapsto t(1+i0^+)$.  In general,
however, this is not possible. To interpret the results of computer
simulations such as quantum Monte Carlo and to make comparisons with
experiment, we require a technique that reliably extracts spectral
information from imaginary time data.  At issue is how best to do this
given that the input data is intrinsically noisy and incomplete.

The most widely used technique is the maximum entropy method
(MEM),~\cite{Gull84,Silver90,Gubernatis91} which selects the best
candidate solution that is consistent with the data.  Here, ``best''
means most likely in the Bayesian sense.  There are several variations
on the algorithm, but in general it plays out as a competition between
the goodness-of-fit measure $\chi^2$ and the entropic prior
$\mathcal{S}$.  In practice, one minimizes the functional $\chi^2 -
\alpha^{-1}\mathcal{S}$ (for some $\alpha^{-1} \neq 0$).  The presence
of the entropic prior introduces a non-linearity that pulls the minimum
away from the least squares solution.  One of the key advantages to the
method is that it is rigourously derived from statistical considerations
and guarantees a unique solution.

Another strategy is to generate a sequence of possible solutions and
then take their mean, with the hope that spurious features will be
averaged out and legitimate features reinforced (as, \emph{e.g.}, in
Ref.~\onlinecite{Mishchenko}).  Such methods, however, tend to be
\emph{ad hoc} and are not rigourously justified. There are no criteria
for selecting which solutions to include or for assigning their relative
weights in the sum.  Moreover, how these schemes are related to the MEM
solution is unclear. There is no reason \emph{a priori} to believe that
an average over several possible spectra will be closer to the true
spectrum than the single most probable one.

Nonetheless, there is compelling evidence that averaging methods can
produce better spectra than the MEM. In particular,
Sandvik~\cite{Sandvik} has shown that an unbiased thermal average of all
possible spectra, Boltzmann weighted according to $\chi^2$, produces (in
several test cases) an average spectrum that is in better agreement with
the true spectrum (found via exact diagonalization) than is the MEM
result. Indeed, our own experience suggests that the MEM is unduly
biased toward smooth solutions: sharp spectral features tend to be
washed out or obliterated.

In this paper, we show how the averaging approach can be made
systematic. We relate the analytic continuation problem to a system of
interacting classical fields living on the unit interval and prove that
the MEM solution is realized as its mean field configuration. From that
point of view, Sandvik's method amounts to allowing thermal fluctuations
about this mean field configuration. It is, in some sense, the most
natural dynamical generalization of the MEM. Finally, we sketch out an
improved algorithm for performing the stochastic sampling and provide
test results for the two methods applied to the spectrum of a simple BCS
superconductor.

\section{Analytic Continuation}

A dynamical correlation function of imaginary time, $G(\tau) = \langle
\T[\hat{O}(\tau)\hat{O}^{\dagger}(0)]\rangle$, satisfies the
(anti-)periodicity relation $G(\tau + \beta) = \mp G(\tau)$, where the
upper sign holds for fermionic operators and the lower sign 
for bosonic ones.  Since it is uniquely determined by its values in the
region $\tau \in [0,\beta)$, the function admits a discrete Fourier
transform
\begin{align} \label{EQ:Gtau}
G(\tau) &= \frac{1}{\beta}\sum_{\omega_n}
e^{-i\omega_n\tau}G(\omega_n),\\ \label{EQ:Gomegan}
G(\omega_n) &= \int_0^\beta \!d\tau\, e^{i\omega_n\tau}G(\tau),
\end{align}
where the sum is over the Matsubara frequencies $\omega_n =
(2n+1)\pi/\beta$ for fermions and $\omega_n =  2n\pi/\beta$ for bosons,
with $n \in \mathbb{Z}$.

Provided that $\lvert G(\omega_n) \rvert$ falls off at least as fast as
$1/\lvert \omega_n \rvert$ when $n \rightarrow \infty$ (which is
guaranteed so long as the operator (anti-)commutator satisfies $\langle
\hat{O}\hat{O}^{\dagger}\pm\hat{O}^{\dagger}\hat{O}\rangle < \infty$),
the Fourier components are representable in terms of a function of the
form
\begin{equation}\label{EQ-Gz}
\mathcal{G}(z) = \mp \int_{-\infty}^\infty \!\frac{d\omega}{2\pi} 
\frac{\rho(\omega)}{z-\omega}
\end{equation}
with the identification $G(\omega_n) = \mathcal{G}(i\omega_n)$.  The
function $\rho(\omega)$ is real-valued and satisfies $\rho(\omega) \ge
0$ for fermions and $\sgn(\omega)\rho(\omega) \ge 0$ for bosons. Note
that $\mathcal{G}(z)$ is analytic everywhere in the complex plane, with
the possible exception of the real line.  Wherever $\rho(\omega)$ is
nonzero, there will be a corresponding jump in $\mathcal{G}(z)$:
\begin{equation}\label{EQ-jump} 
\mathcal{G}(\omega + i0^+) - \mathcal{G}(\omega - i0^+) = \pm
\rho(\omega).
\end{equation}  

The principle of analytic continuation states that given the value of
$\mathcal{G}(z)$ at a countably infinite number of points along the
imaginary axis---by which we mean that $G(\omega_n)$ or, equivalently,
$G(\tau)$ is known---we can \emph{uniquely} extend $\mathcal{G}(z)$ from
those points to the full complex plane. In particular, we can find its
values just above and just below the real axis and hence, via
Eq.~(\ref{EQ-jump}), extract $\rho(\omega)$.

According to Eq.~(\ref{EQ-Gz}), we can write
\begin{equation}
G(\omega_n) = \mp \int \!\frac{d\omega}{2\pi} \frac{\rho(\omega)}
{i\omega_n-\omega}.
\end{equation}
Transforming back to imaginary time, via Eq.~\eqref{EQ:Gtau}, and
performing the Matsubara frequency sum yields
\begin{equation} \label{EQ-lin-func}
\begin{split}
G(\tau) &= \mp \int \!\frac{d\omega}{2\pi}
\frac{1}{\beta}\sum_{\omega_n}
           \frac{e^{-i\omega_n\tau}}{i\omega_n-\omega}\rho(\omega)\\
        &= \int \!\frac{d\omega}{2\pi}
\frac{e^{-\omega\tau}\rho(\omega)}
                {e^{-\beta\omega}\pm 1}\\
        &= \int \!d\omega \, K(\tau,\omega)A(\omega).
\end{split}
\end{equation}
In the last line, we have defined
\begin{equation}
K(\tau,\omega) = \begin{cases}
e^{-\omega\tau}/(e^{-\beta\omega} + 1) & \text{fermions}\\
\omega e^{-\omega\tau}/(e^{-\beta\omega} - 1) & \text{bosons}
\end{cases}
\end{equation}
and
\begin{equation} \label{EQ:Arho}
A(\omega) = \begin{cases}
\rho(\omega)/2\pi & \text{fermions}\\
\rho(\omega)/2\pi\omega & \text{bosons.}
\end{cases}
\end{equation}
(For some applications it may be more appropriate to define
$K(\tau,\omega) = e^{-\omega\tau}$ and $A(\omega) = \rho(\omega)/2\pi
(e^{-\beta\omega}-1)$ in the bosonic case.) The \emph {spectral
function} $A(\omega)$, which we shall view as the main quanitity of
interest, is positive definite and satisfies a sum rule $\int
\!d\omega\, A(\omega) = \mathcal{N} < \infty$. 

Equation~(\ref{EQ-lin-func}) tells us that we can interpret  $G(\tau)$
as a 
linear functional of $A(\omega)$ with kernel $K(\tau,\omega)$. Hence,
the analytic continutation is equivalent to the functional inversion
$A(\omega) = \mathbf{K}^{-1}[G(\tau)]$. Only a finite inversion is
practicable, however.
If we discretize frequency and imaginary time using a uniform mesh (with
spacings 
$\Delta\tau$ and $\Delta\omega$), then $A_j = A(\Delta\omega \cdot
j)\Delta 
\omega$ and $G_k = G(\Delta\tau \cdot k)$ are related by
$A_j = \sum_k K^{-1}_{jk} G_k$. The problem is thus reduced to a matrix 
inversion of 
\begin{equation}
K_{kj} = \frac{e^{\Delta\omega\Delta\tau\cdot j \cdot k}}
{e^{-\beta \Delta \omega \cdot j}\pm 1}.
\end{equation}

This inversion is not an easy one to perform, however.  The condition
number of $K_{jk}$ is extremely large: the matrix will have eigenvalues
both exponentially large and exponentially small in $\beta$.  This means
that computation of the inversion requires extremely high numerical
precision.~\cite{Beach} Worse, the inversion problem is ill-posed and
responds badly to any measurement error in the input set $G_k$. The
inversion typically overfits the noise with spurious high-frequency
modes in $A_j$.

The history of practical analytic continuation methods is one of
continual refinement of the procedures for regularization of the matrix
inversion. The simplest example of regularization is to try
\begin{equation}
A_j = \sum_k (K_{kj}+\lambda\delta_{kj})^{-1} G_k.
\end{equation} 
Since the high-frequency modes in $A_j$ are generated by the smallest
eigenvalues of $K_{jk}$, a nonzero value of $\lambda$ will have the
effect of suppressing those modes with eigenvalues on the order or
$\lambda$ or smaller.  To see this, note that for each eigenvalue $E$ of
$K_{jk}$, there is an eigenvalue in the inverse matrix that is modified
according to $1/E \rightarrow 1/(E + \lambda)$.

This naive scheme has two major flaws. First, filtering out the high
frequency modes in this way has the effect of eliminating from the
spectral function \emph{all} fine structure below a certain frequency
scale, whether spurious or real. Second, it does not ensure that $A_j
\ge 0$, as required.  The MEM, which we describe briefly in the next
section, is considerably more sophisticated about what to filter and has
nonnegativity built in.

\section{\label{SECT:maxentmeth}Maximum Entropy Method}

Suppose that to the exact function $G(\tau)$ we have a measured
approximation $\bar{G}(\tau)$. In practice, this will usually have
been generated from some Monte Carlo simulation, so that
\begin{equation}
\bar{G}(\tau) = G(\tau) + \text{statistical noise.}
\end{equation}

The goodness-of-fit functional
\begin{equation} \label{EQ:chi_squared}
\chi^2[A] = \int_0^\beta \!\frac{d\tau}{\sigma(\tau)^2}\, \left\lvert
\int \!d\omega\, K(\tau,\omega)A(\omega) - \bar{G}(\tau) \right\rvert^2
\end{equation}
measures how closely the correlation function generated from $A(\omega)$
[via Eq.~(\ref{EQ-lin-func}), the forward model] matches
$\bar{G}(\tau)$. Here, $\sigma(\tau)$ is the best-guess estimate of the
total measurement error in $\bar{G}(\tau)$. (See
Appendix~\ref{APP:Discretization}.) There is also an entropy
associated with each spectral function,
\begin{equation} \label{EQ:SA}
\mathcal{S}[A] = -\int \!d\omega\,A(\omega)\ln\left(A(\omega)/D(\omega) 
\right),
\end{equation}
which measures the information content of $A(\omega)$. Here, $D(\omega)$
is the so-called \emph{default model}, a smooth function that serves as
the zero (maximum) entropy configuration. Any features of the true
spectral function known in advance can be encoded in $D(\omega$).

It can be shown that the likelihood of any $A(\omega)$ being the true
spectral function is equal to 
$\mathcal{P}[A] \sim e^{-Q[A]}$ where $Q = \chi^2 -
\alpha^{-1}\mathcal{S}$ (and $\alpha^{-1}$ is a parameter that controls
the degree of regularization). The MEM solution corresponds to the
spectral function that minimizes $Q$. In practice, the minimization of
$Q$ is treated as a numerical optimization problem and is typically
performed using the Newton-Raphson algorithm or some other gradient
search technique. Nonetheless, a formal solution can be found by
identifying the spectral function for which $Q$ is stationary with
respect to functional variation. The result, derived in
Appendix~\ref{APP:formalsol}, is
\begin{equation} \label{EQ:MEM-formal-solution}
\bar{A}(\omega) = e^{\alpha\mu} D(\omega)\exp\biggl[-2\alpha
\int_0^\beta
\!\frac{d\tau}{\sigma(\tau)^2}\,\psi(\tau)K(\tau,\omega)\biggr]
\end{equation}
where
\begin{equation}
\psi(\tau) = \int \!d\omega\,K(\tau,\omega)\bar{A}(\omega) -
\bar{G}(\tau)
\end{equation}
and $\mu$ is a Lagrange multiplier chosen to enforce the normalization
$\int \!d\omega\, A(\omega) = \mathcal{N}$.

In two trivial limits, this set of equations can be solved exactly. When
$\alpha \rightarrow \infty$, Eq.~\eqref{EQ:MEM-formal-solution} demands
that $\psi \rightarrow 0$.  This yields the noisy, unregularized
spectrum $\bar{A}(\omega) = \mathbf{K}^{-1}[\bar{G}(\tau)]$, which is
the solution that minimizes $\chi^2[A]$ . When $\alpha \rightarrow 0$,
$\bar{A}(\omega) = D(\omega)$, the smooth default function. This
solution maximizes $\mathcal{S}[A]$. Note that these results come about
because $Q \sim \chi^2[A]$ and $Q \sim -\mathcal{S}[A]$, respectively,
in the two limits.

Over the full range of intermediate values ($0 < \alpha < \infty$),
Eq.~\eqref{EQ:MEM-formal-solution} constitutes a one-parameter family of
solutions interpolating between these two extremes.  An additional
condition must be imposed to remove this ambiguity, \emph{i.e.}, to turn
the family of solutions into a single final spectrum. In \emph{classic}
MEM, one takes the point of view that somewhere between over-fitting and
over-smoothing lies an ideal intermediate range centred on some optimal
value of $\alpha$. In other schemes, the final result is produced by
averaging, $\bar{A}(\omega) = \int_0^\infty\!d\alpha \, w(\alpha)
\bar{A}(\alpha,\omega) /  \int_0^\infty\!d\alpha \, w(\alpha)$, in which
case the question becomes which weighting function $w(\alpha)$ to use.
In their definitive review,~\cite{Jarrell96} Jarrel and Gubernatis
address these issues in greater detail.

\section{\label{SECT-Alternative-Approach}The Stochastic Approach}

In this section and the next, we introduce the stochastic analytic
continuation approach and demonstrate how it is related to the MEM. To
start, consider a smooth mapping $\phi : \mathbb{R} \mapsto [0,1]$,
which takes the frequency domain of the spectral function onto the unit
interval.  Such a function will be of the form
\begin{equation} \label{EQ:phimap}
\phi(\omega) = \frac{1}{\mathcal{N}}\int_{-\infty}^\omega \!d\nu\,
D(\nu)
\end{equation}
where $D = \mathcal{N}\phi'$ is positive definite and (like $A$)
normalized to $\mathcal{N}$ but otherwise arbitrary. (We use the
notation $D$ for the mapping's kernel in anticipation of identifying it
with the default model of the MEM.) Then,
\begin{equation} \label{EQ:nxnorm}
1 = \frac{1}{\mathcal{N}}\int \!d\omega\, A(\omega) 
                         = \int \!d\phi(\omega)
                            \frac{A(\omega)}{D(\omega)}
                         = \int_0^1 \!dx\, n(x).
\end{equation}
In the last line, we have made the change of variables
$x = \phi(\omega)$ and introduced the dimensionless field
\begin{equation} \label{EQ:n_field_def}
n(x) = \frac{A(\phi^{-1}(x))}{D(\phi^{-1}(x))}
\end{equation}
which, according to Eq.~\eqref{EQ:nxnorm}, is normalized to unity.

Under this change of variables, Eq.~(\ref{EQ:chi_squared}) becomes
\begin{equation} \label{EQ:Hnx}
H[n(x)] = \int_0^\beta \!\frac{d\tau}{\sigma(\tau)^2}\left\lvert
\int_0^1 \!dx\, \hat{K}(\tau,x)n(x) - \bar{G}(\tau) \right\rvert^2
\end{equation}
with $\hat{K}(\tau,\phi(\omega)) = K(\tau,\omega)$. We take the point of
view that Eq.~\eqref{EQ:Hnx} is the Hamiltonian for the system of
classical fields $\{n(x)\}$. Then, supposing the system is held fixed at
a fictitious inverse temperature $\alpha$, it has a partition function
$Z = \int \!\mathcal{D}n\, e^{-\alpha H[n]}$ with a measure of
integration
\begin{equation} \label{EQ:measureint}
\int\!\mathcal{D}n = \int_0^\infty \!\biggl(\prod_x dn(x)\biggr)
\delta\biggl(\int_0^1\!dx\,n(x) - 1 \biggr).
\end{equation}
The thermally averaged value of the field is
\begin{equation} \label{EQ:nalpha}
\langle n(x) \rangle = \frac{1}{Z}\int \!\mathcal{D}n\, n(x) e^{-\alpha
H[n]}.
\end{equation}
The corresponding ``thermally regulated'' spectral function,
\begin{equation}
\langle{A}(\omega)\rangle = \langle n(\phi(\omega)) \rangle D(\omega), 
\end{equation}
can be recovered using Eq.~(\ref{EQ:n_field_def}). 

At zero temperature ($\alpha \rightarrow \infty$), Eq.~\eqref{EQ:nalpha}
simply picks out the ground-state field configuration; the corresponding
spectral function is the unregularized analytic continuation result.  In
the high temperature limit ($\alpha \rightarrow 0$),
Eq.~\eqref{EQ:nalpha} represents an \emph{unweighted} average over all
possible field configurations. In that case, the average is completely
independent of the input function $\bar{G}(\tau)$ and as such can only
yield the zero-information result $\langle n(x) \rangle = 1$. From
Eq.~\eqref{EQ:n_field_def}, it follows that $D(\omega)$ is the
corresponding spectral function.

These limits are precisely those of the MEM, which we discussed at the
end of Sect.~\ref{SECT:maxentmeth}. Note that the kernel of the mapping
in Eq.~\eqref{EQ:phimap} plays the same role as the MEM's default model
and the fictitious temperature the same role as the MEM's regularization
parameter.

\section{\label{SECT:approx}Approximate Solutions}

Now let us extend our ``interacting classical field'' analogy a little
further. Expanding the square in Eq.~\eqref{EQ:Hnx}, we can cast the
Hamiltonian in the familiar form
\begin{multline}
H[n(x)] = \int_0^1 \!dx\, \epsilon(x) n(x)\\ + \frac{1}{2}\int_0^1
\!dx\,dy\,V(x,y)n(x)n(y),
\end{multline}
with a free dispersion
\begin{equation}
\epsilon(x) = -2\int_0^\beta \!\frac{d\tau}{\sigma(\tau)^2}\,
\bar{G}(\tau)\hat{K}(\tau,x)
\end{equation}
and an interaction term
\begin{equation}
V(x,y) = V(y,x) = 2\int_0^\beta
\!\frac{d\tau}{\sigma(\tau)^2}\,\hat{K}(\tau,x)
\hat{K}(\tau,y).
\end{equation}

\emph{Noninteracting system}---Let us ignore the interaction term for a
moment and proceed by setting $V=0$.  Then, if we represent the delta
function constraint in Eq.~\eqref{EQ:measureint} with an integral
representation
\begin{equation}
\delta(X) = \int_{-\infty}^\infty\!d\zeta \, \exp\bigl(i\zeta X \bigr),
\end{equation}
the partition function is simply
$\mathcal{Z} = \int_{-\infty}^\infty\!d\zeta \, e^{-i\zeta}
\mathcal{Z}(\zeta)$, where
\begin{equation}
\mathcal{Z}(\zeta) = \int_0^\infty \!\biggl(\prod_x dn(x)\biggr)
e^{-\int_0^1\!dx\,(\alpha\epsilon(x)
-i\zeta)n(x)}.
\end{equation}
The saddle point solution for the field is
\begin{equation} \label{EQ:nonintsol}
\bar{n}(x) = \frac{\delta}{\delta \epsilon(x)}\biggl(-\frac{1}{\alpha}
\ln Z(\bar{\zeta})\biggr) = e^{-\alpha(\epsilon(x)-\mu)}.
\end{equation}
This says that the fields are Maxwell-Boltzmann distributed according to
their energy as measured with respect to a chemical potential $\mu
\equiv i\bar{\zeta}/\alpha$, which is chosen such that $\int_0^1 \!dx\,
\bar{n}(x) = 1$.

\emph{Mean field treatment}---Now let us reintroduce $V$. Assuming that
fluctuations of the $n(x)$ field about its mean value are negligible,
\begin{equation}
\bigl(n(x) - \bar{n}(x)\bigr) \bigl(n(y) - \bar{n}(y)\bigr) \approx 0,
\end{equation}
the Hamiltonian has a mean field form
\begin{equation} \label{EQ:HMF}
H_{\text{MF}} = \int_0^1 \!dx\, E(x)n(x) + \text{const.},
\end{equation}
where
\begin{equation} \label{EQ:Ex}
E(x) = \frac{\delta H[n]}{\delta n(x)}\biggr\rvert_{n=\bar{n}}
= \epsilon(x) + \int \!dy\, V(x,y)\bar{n}(y).
\end{equation}
Equation~\eqref{EQ:HMF} leads to the saddle point solution given by
Eq.~\eqref{EQ:nonintsol} but now with $\epsilon(x)$ replaced by $E(x)$.
Using the definition of $E(x)$ from Eq.~\eqref{EQ:Ex}, we arrive at the
self-consistent equation
\begin{equation} \label{EQ:nxformal}
\bar{n}(x) = e^{\alpha\mu}\exp\left[-\alpha\Bigl(\epsilon(x)+\int
\!dy\,V(x,y) 
\bar{n}(y) \Bigr)\right].
\end{equation}
Again, $\mu$ is a chemical potential used to fix the normalization.

Now consider the reverse change of variables taking $n(x)$ back to
$A(\omega)$. With only a little effort, one can show that
Eq.~\eqref{EQ:nxformal} is identical to
Eq.~\eqref{EQ:MEM-formal-solution}. What this tells us is that the mean
field treatment of the classical field system is \emph{formally
equivalent to the MEM}.

We can make this equivalence more explicit still. The free energy
density of the system we have just described is $F = U - \alpha^{-1}S -
\mu$, where the internal energy is given by $U = H[\bar{n}(x)]$ and the
entropy (see Appendix~\ref{APP:Entropy}) by 
\begin{equation} \label{EQ:Snx}
S[\bar{n}] = -\int_0^1 \!dx\, \bar{n}(x) \ln \bar{n}(x).
\end{equation}

As we saw earlier, Eqs.~\eqref{EQ:chi_squared}  and \eqref{EQ:Hnx} are
connected by a change of variables. Similarly,
\begin{equation} \label{EQ:Snx}
\begin{split}
S[\bar{n}] &= -\int_0^1 \!dx\, \bar{n}(x) \ln \bar{n}(x)\\
  &= -\int \!d\phi(\omega) \frac{\bar{A}(\omega)}{D(\omega)}\ln\left(
     \frac{\bar{A}(\omega)}{D(\omega)}\right)\\
  &= -\int \!d\omega \, \bar{A}(\omega) \ln\left(
     \frac{\bar{A}(\omega)}{D(\omega)}\right) = \mathcal{S}[\bar{A}],
\end{split}
\end{equation}
where the final equality follows from comparison with Eq.~\eqref{EQ:SA}.
Thus, $\chi^2 = H[\bar{n}(x)]$ and $\mathcal{S} = S[\bar{n}(x)]$, which
makes clear that $F\mathcal{N} = Q = \chi^2 - \alpha^{-1}\mathcal{S} -
\mu \mathcal{N}$. This means that the MEM solution is just the one that
minimizes the free energy of the $\{n(x)\}$ system at the mean field
level.

\section{Monte Carlo Evalutaion}
\subsection{Configurations and Update Scheme}

The energy of a given field configuration, given by Eq.~\eqref{EQ:Hnx},
can be written in the form
\begin{equation} \label{EQ:Hh2}
H[n(x)] = \int_0^\beta \!d\tau\, h(\tau)^2,
\end{equation}
where
\begin{equation}
h(\tau) = \frac{1}{\sigma(\tau)}\int_0^1 \!dx\, \hat{K}(\tau,x)n(x) -
g(\tau)
\end{equation}
and $g(\tau) = \bar{G}(\tau)/\sigma(\tau)$ is the input Green's function
rescaled by the variance.

\begin{figure}
\includegraphics{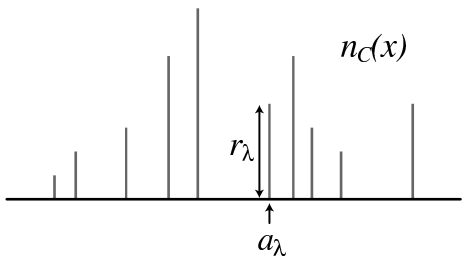}
\caption{A field configuration of delta functions $n_C(x)$ is specified
by a set $C = \{r_\gamma, a_\gamma \}$ of residues and coordinates.}
\end{figure}

Computing $\langle n(x) \rangle$ requires that we integrate over all
possible field configurations. To accomplish this, we need some ansatz
to render the measure $\mathcal{D}n$ finite. One choice is to  represent
each field configuration as a superposition of delta functions. In that
case, we can parameterize each configuration by a set of residues and
coordinates $C = \{r_\gamma, a_\gamma \}$ satisfying $r_\gamma > 0$, $0
\le a_\gamma \le 1$, and $\sum_\gamma r_\gamma = 1$. The corresponding
field configuration is
\begin{equation}
n_C(x) = \sum_\gamma r_\gamma\, \delta(x - a_\gamma ).
\end{equation}
The partition function $\mathcal{Z} = \int dC\, \exp(-\alpha H_C)$ has a
new computationally tractable measure
\begin{equation}
\int\!dC = \prod_\gamma \int_0^\infty\!dr_\gamma
\int_{0}^{1}\!da_\gamma\,\delta\biggl(\sum_\gamma r_\gamma - 1 \biggr).
\end{equation}

In order to calculate the energy $H_C$ of a given configuration via
Eq.~\eqref{EQ:Hh2}, we shall need the relation
\begin{equation}
\begin{split}
g(\tau) + h_C(\tau) &= \frac{1}{\sigma(\tau)}\int_0^1\!dx\,
\hat{K}(\tau,x)n_C(x)\\
&= \frac{1}{\sigma(\tau)}\sum_\gamma r_\gamma \hat{K}(\tau,a_\gamma).
\end{split}
\end{equation}
Now suppose that the configuration is modified ($C \mapsto C'$) by
altering the parameters in some subset $\Lambda$ of the delta function
walkers:
\begin{equation}
\begin{split}
r_\gamma \mapsto r'_\gamma & = r_\gamma + \sum_{\lambda\in\Lambda}
\delta_{\gamma\lambda}\Delta r_\lambda,\\
a_\gamma \mapsto a'_\gamma & = a_\gamma + \sum_{\lambda\in\Lambda}
\delta_{\gamma\lambda} \Delta a_\lambda.
\end{split}
\end{equation}
Accordingly, $h_C \mapsto h_{C'} = h_C + \Delta h$, where
\begin{equation}
\Delta h(\tau) = \frac{1}{\sigma(\tau)}\sum_{\lambda\in\Lambda} \Bigl[
r'_{\lambda}
\hat{K}(\tau,a'_\lambda) - r_{\lambda}\hat{K}(\tau,a_\lambda) \Bigr].
\end{equation}
The configuration energy changes to
\begin{equation} \label{EQ:Delta_H}
\begin{split}
H_{C'} &= \int_0^\beta \!d\tau\, \bigl(h_C(\tau) + \Delta
h(\tau)\bigr)^2\\
&= H_C + \int_0^\beta \!d\tau\,\Delta h(\tau)\bigl[2h_C(\tau) + \Delta
h(\tau)\bigr].
\end{split}
\end{equation}

The Monte Carlo procedure is to calculate $H_C$ and $h_C(\tau)$ for some
arbitrary starting configuation $C$ and then update them whenever a walk
is accepted.  Acceptance is determined according to the usual Metropolis
algorithm: create a modified trial configuration and compute its energy
shift $\Delta H = H_{C'} - H_C$ following Eq.~(\ref{EQ:Delta_H}); choose
a random real number $\xi \in [0,1]$; if $\exp(-\alpha \Delta H) > \xi$,
accept the walk and update
\begin{equation}
\begin{split}
H_C &\mapsto H_{C'} = H_C + \Delta H, \\
h_C & \mapsto h_{C'} = h_C + \Delta h.
\end{split}
\end{equation}

The path of the delta function walkers through the configuration space
must be normalization-conserving and must satisfy detailed balance.
Moreover, the entire phase space must, in principle, be accessible. Only
two types of moves are necessary to meet these criteria: (1) coordinate
shifting moves, in which the walker $\lambda$ is translated by a
distance $\Delta a_\lambda$, and (2) weight sharing moves, in which the
total residue of a subset of walkers is reapportioned amongst themselves
such that $\sum_\gamma r_\gamma = 1$ is preserved.

It is useful, however, to introduce additional weight sharing moves that
also conserve higher moments
\begin{equation}
\mathcal{M}^{(n)} = \int_0^1\!dx\,n(x)x^n = \sum_\gamma r_\gamma
E_\gamma^n.
\end{equation}
Sandvik has shown that such moves dramatically improve the acceptance
ratio of attempted walks at low temperature. At a minimum we want to
consider walks that preserve the overall normalization
$\mathcal{M}^{(0)} = 1$. But we also consider rearrangements of weight
between $n > 2$ walkers that conserve the first $n-1$ moments.  Such a
move can be effected as follows. Let $\Lambda = \{\lambda_1, \lambda_2,
\ldots , \lambda_n\} = \{\lambda_1\} \cup \tilde{\Lambda}$. Defining the
scale factors
\begin{equation}
Q_{\lambda} = \begin{cases}
-1 & \text{if $\lambda = \lambda_1$}\\
\frac{\prod_{\mu \in \tilde{\Lambda}} (a_\mu - a_{\lambda_1}
)}{\prod_{\mu \in \tilde{\Lambda}} (a_\mu - a_\lambda)} & \text{if
$\lambda \in \tilde{\Lambda}$,}
\end{cases}
\end{equation}
we can express the changes in residue as
\begin{equation}
r'_{\lambda} = r_{\lambda} + \Delta r_\lambda = r_{\lambda}
-s\,Q_{\lambda},
\end{equation}
where $s$ parameterizes the 1-dimensional line of constraint through the
$n$-dimensional space of residues. In order to preserve the positivity
of the residues, we must impose $r'_\lambda > 0$. Hence, we need to
ensure that $r_\lambda > Q_\lambda \Delta r_\lambda$ for all $\lambda
\in \Lambda$.  Accordingly, we take $s$ to be randomly distributed in
the interval
\begin{equation}
\max_{\lambda \in \Lambda^-} \bigl(r_\lambda/Q_\lambda\bigr) < s <
\min_{\lambda \in \Lambda^+} \bigl(r_\lambda/Q_\lambda\bigr),
\end{equation}
where $\Lambda^- = \{ \lambda : Q_{\lambda} < 0 \}$ and $\Lambda^+ = \{
\lambda : Q_{\lambda} > 0 \}$.

\subsection{Parallel Tempering}

The Monte Carlo algorithm described above can be improved by introducing
parallel tempering.~\cite{Marinari} The idea is to allow multiple
instantiations of the simulation to proceed simultaneously for a variety
of parameters $\{\alpha_0, \alpha_1, \ldots, \alpha_N\}$ covering a
large range of inverse temperatures. The temperature profile is
arbitrary, but we shall find it convenient to choose a constant ratio
$\alpha_{p+1} / \alpha_p = R$ between one temperature layer and the
next. 

Most important, the field configurations in each layer are made to
evolve in parallel but not independently. Configurations are swapped
between adjacent layers in such a way that preserves detailed balance
and ensures that each layer $p$ will eventually settle into thermal
equilibrium at inverse temperature $\alpha_p$. The update rule is quite
simple: given two adjacent layers $p$ and $q=p\pm 1$, choose a random
real number $\xi \in [0,1]$ and swap the $p$ and $q$ configurations if
\begin{equation}
\exp\Bigl[\bigl(\alpha_p-\alpha_q\bigr)
                  \bigl(H_p-H_q\bigr)\Bigr] > \xi.
\end{equation}

\begin{figure}
\includegraphics{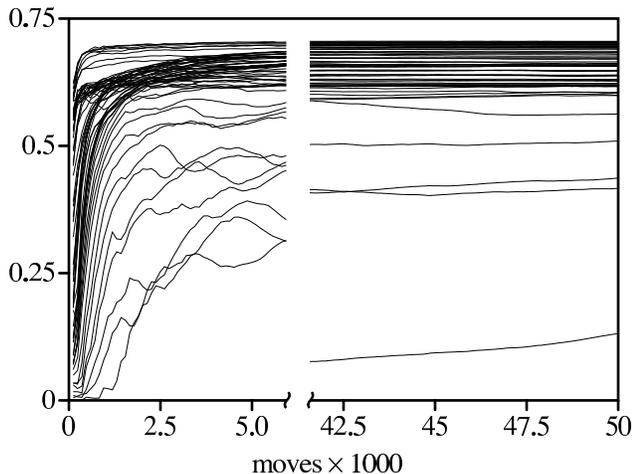}
\caption{\label{FIG:swap}The acceptance ratio of configuration swaps
between adjacent levels ($\alpha_p \leftrightarrow \alpha_{p+1}$)
evolves as a function of the number of updates performed. When the
system is fully thermalized, the acceptance ratios stabilize to
asymptotic values.}
\end{figure}

Parallel tempering eliminates the need for a separate, initial annealing
stage.~\cite{Sandvik} Because the simulation simultaneously samples over
a large temperature range, there is no danger of getting trapped in
false minima: the interlayer walks always provide a cheap pathway
between configurations separated by large energy barriers. All that is
required is to let the system thermalize for some time before sampling
(\emph{i.e.}, before actually beginning to bin and tabulate the field
configurations). By tracking the average acceptance rates for swaps
between layers, it is straightforward to determine when the system has
equilibrated. Figure~\ref{FIG:swap} shows a sample run (for a test case
to be described in Sect.~\ref{SECT:bcs}). We see that on a stochastic
time-scale of several tens of thousands of moves, each temperature layer
settles into thermal equilibrium.

An additional advantage of the parallel tempering algorithm is that it
yields in one run a complete temperature profile of all the important
thermodynamic variables.  In the next section, we discuss how we can put
that information to use.

\section{Critical Temperature}

The Monte Carlo simulation yields a set of thermally averaged field
configurations $\{ \langle n(x) \rangle_{\alpha_p} : p = 0,1,\ldots
N\}$. With little additional effort, we can also keep track of the
internal energies $\{U(\alpha_p) = \langle H[n] \rangle_{\alpha_p} : p =
0,1,\ldots N\}$. In this section, we propose a final candidate spectral
function constructed from only these quantities.

To start, note that the specific heat can be written as
\begin{equation}
C(\alpha_p) = \frac{dU(\alpha)}{d(\alpha^{-1})} \biggr\rvert_{\alpha =
\alpha_p}
\approx \frac{a_p U(\alpha_p)}{\ln R} \frac{d\ln U(\alpha_p)}{dp}.
\end{equation}
(See Appendix~\ref{APP:logmesh}.) In Fig.~\ref{FIG:knee}, $\ln
U(\alpha_p)$ is plotted for each temperature level. The knee in the
function, occurring in the vicinity of the level $p = p^*$, indicates
there is a jump in the specific heat. At low temperatures ($\alpha >
\alpha^* \equiv \alpha_{p*}$), the system freezes out and the
correlations $\langle n(x)n(y) \rangle - \langle n(x) \rangle \langle
n(y) \rangle$ become short-ranged.  There is a characteristic energy
scale $E^* = U(\alpha^*)$ associated with this phase transition.

\begin{figure}
\includegraphics{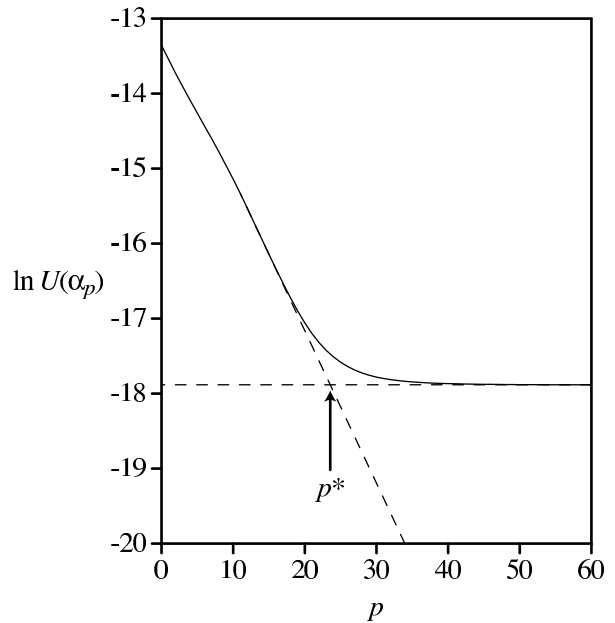}
\caption{\label{FIG:knee}The internal energy of the $\{n(x)\}$ system at
each temperature layer is plotted. The knee at $p = p^*$, corresponding
to a jump in the specific heat, signals a thermodynamic phase
transition.}
\end{figure}

Recall that in the \emph{microcanonical ensemble}, the average over all
configurations having energy $E$ is given by
\begin{equation}
\langle n(x) \rangle_E = \int \mathcal{D}n\, n(x)\delta(E-H[n]).
\end{equation}
We propose that the final spectrum be defined as
\begin{equation} \label{EQ:nfinal}
\langle\langle n(x) \rangle\rangle = \frac{1}{E^*}\int_0^{E^*}\! dE\,
\langle n(x) \rangle_E,
\end{equation}
which sums over all field configurations in the ordered phase
(\emph{i.e.}, configurations with energies $E$ satisfying $0 \le E <
E^*$). Roughly speaking, this amounts to performing an unbiased average
over all spectral functions $A$ that surpass the fitting threshold
$\chi^2[A] < E^*$.

Since the Monte Carlo simulation is performed at fixed temperature,
however, we must make the change of variables $dE =
(dU/d\alpha)d\alpha$. Equation~\eqref{EQ:nfinal} becomes
\begin{equation}
\langle\langle n(x) \rangle\rangle =
\frac{1}{U(\alpha^*)}\int_{\alpha^*}^{\infty}\! d\alpha\,
\biggl(-\frac{dU}{d\alpha}\biggr)\langle n(x) \rangle_{\alpha}.
\end{equation}
The discretized version of this integral is
\begin{equation} \label{EQ:finaln}
\langle\langle n(x) \rangle\rangle = \frac{\sum_{p=p^*}^{N-1}
\bigl(U(\alpha_p) - U(\alpha_{p+1})\bigr)\langle n(x)
\rangle_{\alpha_p}}{U(\alpha_{p^*})-U(\alpha_N)}.
\end{equation}

\section{\label{SECT:bcs}BCS Test Case}

\begin{figure}
\includegraphics{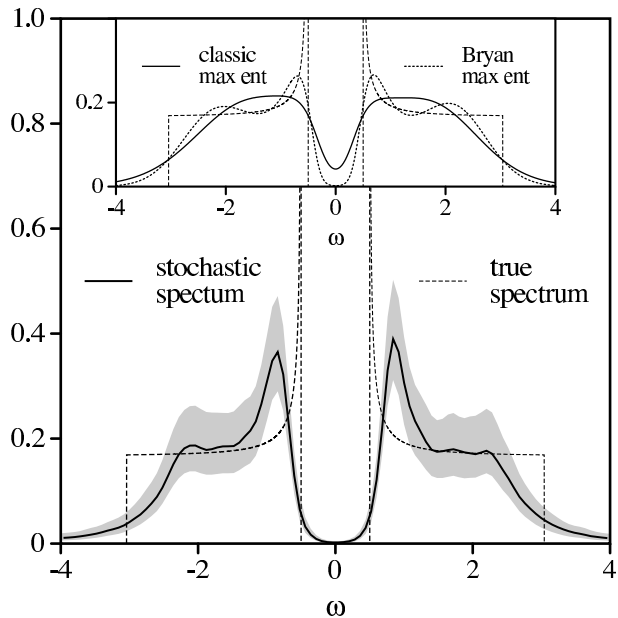}
\caption{\label{FIG:BCS}The stochastic analytic continuation method is
used to extract the spectrum of a BCS superconductor (bandwidth $W=6$
and gap $2\Delta = 1$) from noisy data.  The grey region indicates the
statistical uncertainty of the computed spectrum. The inset shows the
classic and Bryan MEM results.}
\end{figure}

We showed in Sect.~\ref{SECT:approx} that the stochastic analytic
continuation method is a dynamical generalization of the MEM. The
question remains, What is gained by going beyond the mean field
calculation? Our contention is that the stochastic method is better able
to resolve sharp spectral features buried in noisy data. To illustrate
this point, we have taken the spectrum of a BCS superconductor---which
contains flat regions, steep peaks, and sharp gap edges---as a test
case. The exact spectral function is
\begin{equation}\label{EQ:Abcs}
A(\omega) = \begin{cases}
\frac{1}{W}\frac{\lvert \omega \rvert}{\sqrt{\omega^2 - \Delta^2}} &
\text{if $\Delta < |\omega| < W/2$}\\
0 & \text{otherwise},
\end{cases}
\end{equation}
where $W$ is the bandwidth and $2\Delta$ the gap magnitude.

From Eq.~\eqref{EQ:Abcs} we generated an exact $G(\tau)$ using the
forward model. We then applied random error to the function to create an
approximate $\bar{G}(\tau)$, which was made to serve as the input data
for our stochastic algorithm and for two flavours of the MEM---the
classic method and a method due to Bryan\cite{Bryan90} (both described
in Ref.~\onlinecite{Jarrell96}). Figure~\ref{FIG:BCS} shows these
computed spectra alongside the exact result.

The most striking aspect of the comparison is that the stochastically
generated spectrum does a superior job of modelling the gap. It closely
follows the trough of the gap and captures some of the sharpness of the
peaks at the gap edges. The MEM spectra, on the other hand, are much too
smooth. The classic MEM spectrum is especially poor.  It is at best a
caricature of the true BCS spectrum: the sharp features are completely
washed out and the depression at $\omega = 0$ is not a fully developed
gap.

Bryan's algorithm does a somewhat better job of reproducing the gap and
its adjacent peaks, but in doing so it also forms a second pair of
spurious humps around $\omega = 2$. In our experience, this is typical
behaviour. The MEM method has trouble making sudden transitions from
regions of high to low curvature. What one tends to get is a smooth
curve gently oscillating around the correct result. The stochastic
method, in contrast, seems to have no trouble generating a flat region
next to a sharp peak.

\section{Conclusions}

In this paper, we have made the case that the MEM is not the best method
for extracting spectral information from imaginary time data. Instead,
we advocate the use of the stochastic analytic continuation method. Our
claim is that the stochastic method is at least as good as the MEM and
may even surpass it for a broad class of problems in which the spectrum
to be extracted has very sharp features.

This is a difficult point to argue convincingly. New analytic
continuation methods tend to face considerable resistance, and claims of
superiority on their behalf are met (quite rightly) with a high degree
of skepticism. The MEM has a record of years of successful use in a
variety of settings; plus, it offers the comfort of a seemingly
rock-solid mathematical rationale. Competing schemes tend to lack any
clear justification other than a few tantalizing examples of their
performance in a handful of test cases. 

The prevailing opinion is that the MEM is the definitive ``solution'' to
the analytic continuation problem. Some other method may produce better
spectra in particular special cases, but as a general method, the MEM
has to win out. The thinking goes: no other algorithm can outperform the
MEM because its solution is, by construction, the unique, best candidate
spectrum---a claim that rests on the firm foundation of Bayesian logic.

What this line of reasoning misses, however, is the possibility that an
average of \emph{many} likely candidates might better reproduce the true
spectral function than does the \emph{single} most likely spectrum. To
give a path integral analogy, we would argue that including fluctuations
about a saddle point solution (the single most likely field
configuration) can yield a result closer to the full integral. This is
how we go about justifying the stochastic analytic continuation method. 

Let us be careful about what can be established rigourously. To be
precise, the standard conditional probability analysis used to derive
the MEM proves only that the most likely spectrum belongs to the family
of solutions (parameterized by $\alpha^{-1}$) that minimizes $Q = \chi^2
- \alpha^{-1} \mathcal{S}$. From our point of view, then, what is
required of an averaging method is that it produce at the mean field
level a family of solutions that coincides with the MEM result. The
stochastic method, as we have formulated it, does exactly this---under
the guise of minimizing the free energy $F\mathcal{N}$~$(= Q)$ of  a
system of classical fields at a fictitious temperature $\alpha^{-1}$.

This correspondence gives us a new way of thinking about the MEM
solution. We know that even though a path integral contains jagged,
discontinuous field configurations, its saddle point solution is always
a smooth, continuous function. This highlights the main deficiency of
the MEM---that it fails to model well spectral functions that are not
sufficiently smooth---and makes clear why the stochastic method does not
suffer from the same limitation.

Another advantage of the stochastic approach is that it helps us to talk
about the analytic continuation problem using a more physical language.
Having identified the regularization parameter as a temperature, we can
ask how the system behaves thermodynamically. The answer, we have
suggested, is that the system exhibits ordered and disordered phases
that can be interpreted as the good-fitting and ill-fitting regimes. We
believe that this gives a much more intuitive picture than does the
somewhat obscure probability analysis of the MEM.

We close with a recapitulation of the main results. We have presented a
new variant of the stochastic analytic continuation method that differs
from Sandvik's original prescription as follows: as a matter of
mathematical formulation, it includes an additional internal freedom
that turns out to be equivalent to specifying a default model; as a
matter of practical implementation, it is built on a delta function
walker scheme and takes advantage of parallel tempering. We have proved
that the mean field version of this stochastic method is equivalent to
the MEM. Our tests suggest that it outperforms the MEM for spectra with
sharp features and fine structure.

\acknowledgements

The author would like to thank Anders Sandvik for several helpful
discussions and Philippe Monthoux for generously making his maximum
entropy code available. Support for this work was provided by the
Department of Energy under grant DE-FG02-03ER46076.

\appendix

\section{\label{APP:Discretization}Statistical Error and Discretization}

In Eq.~\eqref{EQ:chi_squared}, we have used notational shorthand to
gloss over two subtle issues. First, we have ignored the fact that the
statistical errors between $\bar{G}(\tau)$ and $\bar{G}(\tau')$ are not
independent for $\tau\neq \tau'$. In general, the errors will be
positively correlated whenever $\lvert \tau - \tau' \rvert$ is
sufficiently small. There is also a tendency for them to be negatively
(positively) correlated over long-separated times since $G(0^-) = \mp
G(\beta)$ is built in to the definition of the correlation function.
Thus, one should more properly write the goodness-of-fit measure as
\begin{equation} \label{EQ:chi-cov}
\chi^2[A] = \frac{1}{\beta^2} \!\int d\tau d\tau'\, \Delta(\tau)
C^{-1}(\tau,\tau') \Delta(\tau'),
\end{equation}
where $\Delta(\tau) = \int d\omega\, K(\tau,\omega)A(\omega) -
\bar{G}(\tau)$ and $C(\tau,\tau')$ is the covariance function for
$\bar{G}(\tau)$.

Second, we have ignored the discrete nature of the known input data. A
Quantum Monte Carlo algorithm, for example, is used to generate
stochastically a sequence of independent measurements $\{ G^{(1)},
G^{(2)}, \ldots, G^{(M)} \}$, where each $G^{(m)}$ is an
$(L\!\!+\!\!1)$-vector holding the values of the single-particle
propagator at imaginary times $\tau_l = \beta l/L$ for $l = 0, 1,
\ldots, L$.

The numerical measurement of the Green's function is accomplished by
taking the average
\begin{equation} \label{EQ:Gbar-sumM}
\begin{split}
\overline{G}_l &= \frac{1}{M}\sum_{m=1}^M G^{(m)}_l.\\
\overline{G_l G_{l'}} &= \frac{1}{M}\sum_{m=1}^M G^{(m)}_l G^{(m)}_{l'}.
\end{split}
\end{equation}
The corresponding covariance matrix is given by
\begin{equation} \label{EQ:Cov-sumM}
\begin{split}
C_{ll'} &= \frac{1}{M\bigl(M-1\bigr)}\sum_{m=1}^M \bigl(G^{(m)}_l -
\bar{G}_l \bigr)\bigl(G^{(m)}_{l'} - \bar{G}_{l'} \bigr).\\
&= \frac{1}{M-1}\sum_{m=1}^M \biggl[ \overline{G_l G_{l'}} - \bar{G}_l
\bar{G}_{l'}  \biggr]
\end{split}
\end{equation}
Equation~\eqref{EQ:chi-cov} must now be discretized in order to make use
of Eqs.~\eqref{EQ:Gbar-sumM} and \eqref{EQ:Cov-sumM}. The imaginary time
integrals are carried out numerically on a uniform mesh of $L$ time
slices (spaced by $\Delta\tau = \beta / L$) according to the formula
\begin{equation}
\int_0^\beta d\tau f(\tau) \approx \sum_{l=0}^L \Delta\tau \, w_l \,
f_l,
\end{equation}
where $f_l = f(\Delta\tau\cdot l)$ and the Bode's rule weights $w_l$
satisfy $\sum_{l=0}^L w_l = L$. Equaton~\eqref{EQ:chi-cov} becomes
\begin{equation} \label{EQ:chi-cov-discrete}
\chi^2 = \frac{1}{L^2} \sum_{l,l'=0}^L w_l \Delta_l C^{-1}_{ll'} w_{l'}
\Delta_{l'}.
\end{equation}
Since $\Delta(0) = \pm \Delta(\beta)$,
\begin{equation} \label{EQ:chi-cov-discrete}
\chi^2 = \frac{1}{L^2} \sum_{l,l'=0}^{L-1} \tilde{w}_l \Delta_l
C^{-1}_{ll'} \tilde{w}_{l'} \Delta_{l'}.
\end{equation}
Here, $\tilde{w}_l = w_l + \delta_{l,0} w_L$ for $l = 0, 1, \ldots L-1$.

We now want to solve for the unitary transformation $U$ that
diagonalizes the covariance matrix.  This allows us to write $C =
U^t\Sigma U$ in terms of a set of  \emph{statistically independent}
variances $\Sigma = \diag (\sigma^2_1, \sigma^2_2, \ldots, \sigma^2_L)$.
The inverse matrix is $C^{-1} = U^t \Sigma^{-1} U$.

Putting $C^{-1}_{ll'} = \sum_{k=0}^L \frac{1}{\sigma^2_k} U_{kl}
U_{kl'}$ into Eq.~\eqref{EQ:chi-cov-discrete} yields
\begin{equation}
\begin{split}
\chi^2 &= \sum_{k=0}^L \frac{1}{\sigma^2_k}  \biggl( \frac{1}{L}
\sum_{l=0}^L U_{kl} w_l \Delta_l \biggr)^2\\
&= \sum_{k=0}^L \frac{1}{\sigma^2_k} \Bigl\lvert (VK)[A]_k - (VG)_k
\Bigr\rvert^2
\end{split}
\end{equation}
where we have defined the matrix $V_{kl} = U_{kl} w_l / L$.

To recapitulate, the discretization of the $\tau$ integration is
implicit in Eq.~\eqref{EQ:chi_squared}; it also presumes that we are
working in the $V$ basis in which the covariance matrix is diagonal.

\section{\label{APP:formalsol}Maximum Entropy Formal Solution}

We want to examine the changes in $\mathcal{S}$ with variations in
$A(\omega)$.
Since the spectral function is subject the the normalization
constraint $\int d\omega\, A(\omega) = \mathcal{N}$, variations
in $A(x)$ and $A(y)$ for $x\neq y$ are not independent. We can enforce
the constraint by introducing a lagrange multiplier $\Gamma = 1 +
\alpha\mu$.  Let us define
\begin{equation} \label{EQ:extendS}
\mathcal{S}[A,\Gamma] \equiv -\int d\omega\,A\ln\left(A/D\right)
       +\Gamma\int d\omega\,\left(A-D\right).
\end{equation}
We have assumed here that $D(\omega)$ and $A(\omega)$ have the same
normalization.

Variations of the extended functional, Eq.~\eqref{EQ:extendS}, look like
\begin{equation}
\begin{split}
\frac{\delta \mathcal{S}}{\delta A(x)} &=
-\ln\left(\frac{A(x)}{D(x)}\right) +
\alpha \mu\\
\frac{d \mathcal{S}}{d\mu} &= \alpha\int\!
d\omega\,\left(A(\omega)-D(\omega)\right)
\end{split}
\end{equation}
There is a unique solution that causes these two equations to vanish:
$A(x) = D(x)$, $\mu = 0$.  This implies that $\mathcal{S} = 0$ and
$\delta \mathcal{S} = 0$.

Also, since
\begin{equation}
\frac{\delta^2 \mathcal{S}}{\delta A(x)\delta A(y)} =
-\frac{\delta(x-y)}{A(x)}
\end{equation}
we find that $\delta^2 \mathcal{S} \le 0$.  This means that the entropy
functional
is strictly non-positive and takes its maximum $\mathcal{S}=0$ when $A$
is equal
to the default model.

Similar considerations for $\chi^2[A]$ allow us to construct the total
variation in $Q = \chi^2 - \alpha^{-1} \mathcal{S}$. We find that
\begin{equation} 
\begin{split}
0 = \frac{\delta Q[A,\mu]}{\delta A(\omega)} &= 2\int_0^\beta d\tau\,
K(\tau,x)\psi(\tau)\\
&\quad - \alpha^{-1} \left[- \ln \left( \frac{A(\omega)}{D(\omega)} 
\right) + \alpha\mu\right]
\end{split}
\end{equation}
where
\begin{equation}
\psi(\tau) = \int d\nu\,K(\tau,\nu)A(\nu) - \bar{G}(\tau).
\end{equation}

\section{\label{APP:Entropy}Configurational Entropy}

Consider a system of $N$ energy levels with degeneracies
$m_p$ ($p=1,2,\ldots,N$).  Suppose that each level
is filled with $n_p$ indistinguishable particles.
The state of the system is unchanged by the rearrangement
of particles within a given level.  Thus, given a set of
occupancies $\{0 \le n_p \le m_p\}$, the number of
equivalent configurations is $\Omega(\{n_p\}) = \prod_p\begin{pmatrix}
m_p\\
n_p
\end{pmatrix}$
and the entropy due to this configuration is
\begin{equation}
\ln \Omega(\{n_p\}) = \frac{1}{N}\sum_p\ln 
\begin{pmatrix}
m_p\\
n_p
\end{pmatrix}.
\end{equation}

The binomial coefficient
$\begin{pmatrix}
m\\
n
\end{pmatrix}
= m!/(m-n)!/n!$
can be approximated using 
Stirling's formula $m! \approx m\ln m$.
In the limit of small relative occupancy, this gives
\begin{equation}
\begin{split}
\ln 
\begin{pmatrix}
m\\
n
\end{pmatrix} &= m\ln m - (m-n)\ln (m-n) - n \ln n\\
&\xrightarrow{m>>n} - n \ln n.
\end{split}
\end{equation}

Going over to the continuum, we make the identification
\begin{equation}
\begin{split}
\frac{1}{N}\sum_p &\rightarrow \int \!dx\\
m_p &\rightarrow \infty\\
n_p &\rightarrow n(x)
\end{split}
\end{equation}
and use the counting arguments above to write the entropy
associated with each field configuration:
\begin{equation}
\ln \Omega[n] = -\int_0^1 \!dx\, n(x)\ln n(x).
\end{equation}
The total entropy is
\begin{equation}
S = \int\!\mathcal{D}n\,\ln \Omega[n] \approx \ln \Omega[\bar{n}].
\end{equation}

\section{\label{APP:logmesh}Discretization over a logarithmic mesh}

Suppose that we want to integrate a function $f(\alpha)$ known only at
the points $\alpha_n = R^n \alpha_0$ for $n = 0, 1, \ldots N$. The
integral identity
\begin{equation}
\int \! d\alpha\, f(\alpha) = \int \! d\tilde{\alpha}\,
e^{\tilde{\alpha}} f(e^{\tilde{\alpha}})
\end{equation}
follows from the change of variables $\alpha = \exp(\tilde{\alpha})$. In
this basis, the known points describe a  uniform mesh
\begin{equation}
\tilde{\alpha}_n = \ln \alpha_n = \ln \alpha_0 + n \ln R
\end{equation}
with spacing $\Delta\tilde{\alpha} = \tilde{\alpha}_{n+1} -
\tilde{\alpha}_{n} = \ln R$. Accordingly,
\begin{equation}
\begin{split}
\int \! d\alpha\, f(\alpha) &\approx \sum_{n=0}^N \Delta\tilde{\alpha}\,
e^{\tilde{\alpha_n}} f(e^{\tilde{\alpha}_n})\\
&= \sum_{n=0}^N \bigl(\ln R\bigr) \alpha_n f(\alpha_n).
\end{split}
\end{equation}

When the integrand is of the form
\begin{equation}
\frac{dU}{d\alpha} = \frac{1}{e^{\tilde{\alpha}}}
\frac{dU}{d\tilde{\alpha}}
\end{equation}
we must first discretize the derivative
\begin{equation}
\begin{split}
\frac{dU}{d\alpha}\biggr\rvert_{\alpha = \alpha_n} &\approx
\frac{1}{e^{\tilde{\alpha}_n}} \frac{U(\alpha_{n+1}) -
U(\alpha_n)}{\Delta\tilde{\alpha}}\\
&= \frac{U(\alpha_{n+1}) - U(\alpha_n)}{\alpha_n \ln R},
\end{split}
\end{equation}
which leads to the integrals
\begin{multline}
\int_{\alpha_{p}}^{\alpha_{N-1}}\! d\alpha\,
\biggl(-\frac{dU}{d\alpha}\biggr) \langle n(x) \rangle_{\alpha}\\
\approx \sum_{n=p}^{N-1} \bigl[U(\alpha_n)-U(\alpha_{n+1})\bigr]\langle
n(x) \rangle_{\alpha_n}
\end{multline}
and
\begin{equation}
\begin{split}
\int_{\alpha_{p}}^{\alpha_{N-1}}\! d\alpha\,
\biggl(-\frac{dU}{d\alpha}\biggr) 
&\approx \sum_{n=p}^{N-1} \bigl[U(\alpha_n)-U(\alpha_{n+1})\bigr]\\
&= U(\alpha_p) - U(\alpha_N).
\end{split}
\end{equation}
Equation~\eqref{EQ:finaln} is simply the ratio of these two results.

\bibliography{cont}

\end{document}